\documentclass[9pt,twocolumn,twoside]{opticajnl}

\usepackage{natbib}

\journal{josaa} 

\setboolean{shortarticle}{true}
\usepackage{lineno}
\usepackage{graphicx}
\usepackage{caption}
\usepackage[figurename=Fig.]{caption}
\newcommand{\D}{{\rm d}}

\title{Cross-spectral purity: a generalization of spatiotemporal separability}

\author[1,*]{Matias Koivurova}
\author[2]{Rajneesh Joshi}
\affil[1]{Center for Photonics Sciences, Department of Physics and Mathematics, University of Eastern Finland, P.O. Box 111, 80101, Joensuu, Finland}
\affil[2]{Department of Physics, Government Degree College Danya, 263622, Almora, Uttarakhand, India}

\affil[*]{Corresponding author: matias.koivurova@uef.fi}

\begin{abstract}
We study the connection between cross-spectral purity and spatiotemporal separability of nonstationary (pulsed) scalar fields. It is found that in the case of complete coherence, there is a two-way relation between global cross-spectral purity and spatiotemporal separability of the field. {Moreover, we show that cross-spectral purity generalizes the notion of spatiotemporal separability to partially coherent fields, due to the separability of the correlation functions.} We also briefly discuss simple self-referencing linear measurement methods that can reveal cross-spectral purity.
\end{abstract}
\setboolean{displaycopyright}{true}

\begin{document}
\maketitle

In theoretical investigations spatiotemporal separability is often assumed. That is, one writes the field as a product of temporal and spatial terms. Although it is clear that this is not a general property of light, it has been the norm dating back to the mathematical foundations of photonics. This is mainly due to the fact that it is mathematically convenient. Moreover, fields with significant spatiotemporal coupling have not been possible until relatively recently. With the advent of ultrashort pulses, researchers quickly realized the importance of spatiotemporal coupling, which became an independent field of study \cite{Roadmap}. Today, there are multiple linear and nonlinear measurement methods to establish whether a field is spatiotemporally coupled \cite{Alonso:10,Miranda:14,Pariente2016,Borot:18,Jolly_2020}.

Meanwhile, a significant concept in coherence research called cross-spectral purity, was introduced by Leonard Mandel in the context of stationary scalar light fields in 1961 \cite{mandel1961concept, mandel1995optical}. Generally, when two optical fields interfere, the normalized spectrum at the plane of interference is not a simple mean of the normalized spectra of the interfering fields. If the two interfering fields have identical normalized spectra, and a position exists at the plane of interference where the normalized spectrum is also the same, the field is called cross-spectrally pure \cite{mandel1961concept}. Such fields exhibit an interesting reduction property in their space-time correlations, which is reminiscent of spatiotemporal separability. In fact, this property is often intrinsically assumed, which was pointed out in the context of the Hanbury Brown--Twiss experiment \cite{james1997cross}.

It was shown early on that the space-frequency correlation function of cross-spectrally pure fields is independent of the frequency of the input spectra of light \cite{mandel1976spectral}. Cross-spectral purity has been studied in various scenarios, such as squeezed light, three-dimensional fields, ghost imaging, and scattering \cite{ou1987coherence, james1997cross, liu2007cross, lahiri2014effect}. In recent years, cross-spectral purity has been extended to vector (or electromagnetic) fields \cite{hassinen2009cross, hassinen2011cross, lahiri2013concept, hassinen2013purity, chen2014cross, peng2017cross}, as well as nonstationary scalar \cite{koivurova2019cross, laatikainen2023reduction} and nonstationary electromagnetic light \cite{joshi2021cross, laatikainen2023cross, joshi2024cross}. Importantly, it was shown that cross-spectral purity is generally not preserved upon propagation if the field has any finite bandwidth \cite{koivurova2019cross}. Also, a few techniques have been discussed for generating cross-spectrally pure light \cite{kandpal2002generation, koivurova2019cross, laatikainen2022spectral, joshi2023generation}. 

In the present letter, we examine the conditions for cross-spectral purity and find an intimate connection with spatiotemporal separability. We discuss the cases of complete coherence, partial coherence, and complete incoherence in detail. It is found that global cross-spectral purity ensures that a completely coherent scalar field is spatiotemporally separable. {Further, we show that the notion of cross-spectral purity generalizes the concept of spatiotemporal separability of fields to the spatiotemporal separability of correlation functions.}

We describe the scalar electric field with a complex analytic signal, so that the spectral field is related to the time-domain field through a Fourier transform pair, as in
\begin{subequations}
    \begin{align}
    \label{Fourier_a}
    \tilde{E}(\boldsymbol{\rho};t) = \int_0^{\infty}E(\boldsymbol{\rho};\omega)\exp(-i\omega t)\D\omega,
    \end{align}
    \begin{align}
    \label{Fourier_b}
    E(\boldsymbol{\rho};\omega) = \frac{1}{2\pi}\int_{-\infty}^{\infty}\tilde{E}(\boldsymbol{\rho};t)\exp(i\omega t)\D t,
\end{align}
\end{subequations}
where $\boldsymbol{\rho}$ contains the transverse spatial coordinates $(x,y)$, $t$ is time, and $\omega$ is the angular frequency. Throughout the present work, we denote temporal domain quantities with a tilde when the notation would be otherwise unclear. The complex spectral field can also be written as a product of a real valued envelope and a spectral phase term, as in
\begin{align}
    E(\boldsymbol{\rho};\omega) = A(\boldsymbol{\rho};\omega)\exp[i\varphi(\boldsymbol{\rho};\omega)].
\end{align}
In general, the spectral phase fluctuates around a well-defined mean value, which is the wavefront of the field at the frequency $\omega$. Additionally, we employ the cross-spectral density (CSD), defined as
\begin{align}
\label{eq39}
    W(\boldsymbol{\rho}_1,\boldsymbol{\rho}_2; \omega_1,\omega_2) = \langle E^*(\boldsymbol{\rho}_1;\omega_1) E(\boldsymbol{\rho}_2; \omega_2) \rangle,
\end{align}
where the angle brackets denote ensemble averaging, and the average energy spectrum is given by $S(\boldsymbol{\rho}; \omega) = \langle\left|E(\boldsymbol{\rho}; \omega)\right|^2\rangle = W(\boldsymbol{\rho},\boldsymbol{\rho}; \omega,\omega)$. One can use the average spectrum to define the complex degree of coherence by normalizing as in
\begin{align}
\label{eq75a}
    \mu(\boldsymbol{\rho}_1,\boldsymbol{\rho}_2; \omega_1,\omega_2) = \frac{W(\boldsymbol{\rho}_1,\boldsymbol{\rho}_2; \omega_1,\omega_2)}{\sqrt{S(\boldsymbol{\rho}_1;\omega_1)S(\boldsymbol{\rho}_2; \omega_2)}}.
\end{align}
Note that due to the Fourier transform relation of Eqs.~(\ref{Fourier_a}) and (\ref{Fourier_b}), the following spectral considerations can be straightforwardly translated into the temporal domain. Additionally, one deals with nonstationary fields when $\left|\mu(\boldsymbol{\rho},\boldsymbol{\rho}; \omega_1,\omega_2)\right| \in (0,1]$, and stationary fields when $\left|\mu(\boldsymbol{\rho},\boldsymbol{\rho}; \omega_1,\omega_2)\right| = 0$ for all combinations of $\omega_1$ and $\omega_2$ \cite{koivurova2024nonstationary}. The emphasis of this article will be on scalar nonstationary fields (such as linearly polarized pulsed beams). 

Since we will be discussing spatiotemporal separability, it is important to first establish what it is that separates. In the general partially coherent case, the correlation function may be spatiotemporally separable. This leads to the separability of the average spectrum as well as the average intensity. However, this does not mean that every electric field realization of a given ensemble is separable, but rather that their averages are (although the individual realizations can be separable, in principle). As a particular case, if one has a completely coherent field, for which $\left|\mu(\boldsymbol{\rho}_1,\boldsymbol{\rho}_2; \omega_1,\omega_2)\right| = 1$ at all possible coordinate combinations, then with the use of Eqs.~(\ref{eq39}) and (\ref{eq75a}), as well as the definition of the average spectrum, we see that $\left|\langle E^*(\boldsymbol{\rho}_1;\omega_1) E(\boldsymbol{\rho}_2; \omega_2) \rangle\right| = \langle\left|E(\boldsymbol{\rho}_1; \omega_1)\right|\rangle \langle\left|E(\boldsymbol{\rho}_2; \omega_2)\right|\rangle$. In other words, there are no amplitude or phase fluctuations in the ensemble of realizations, and in this special case the spatiotemporal separability of the correlation function directly corresponds to the spatiotemporal separability of the field.

The condition of cross-spectral purity that Mandel set forth is that the normalized spectra belonging to two different spatial positions of a field are equal to each other, as well as the spectrum at a plane where they are superposed. This condition can be mathematically expressed as
\begin{align}
\label{condition}
    S(\boldsymbol{\rho}_1; \omega) 
    = \frac{S(\boldsymbol{\rho}_2;\omega)}{C(\boldsymbol{\rho}_1,\boldsymbol{\rho}_2)}
    = \frac{S(\boldsymbol{R};\omega)}{D(\boldsymbol{R},\boldsymbol{\rho}_1)},    
\end{align}
where $\boldsymbol{R}$ is the point of observation at the plane of interference, and $C(\boldsymbol{\rho}_1,\boldsymbol{\rho}_2)$ and $D(\boldsymbol{R},\boldsymbol{\rho}_1)$
are real valued, position dependent scaling factors, which ensure that the spectra at $\boldsymbol{\rho}_2$ and $\boldsymbol{R}$ are scaled to coincide with the spectrum at $\boldsymbol{\rho}_1$. Note that one can consider either \textit{local} or \textit{global} cross-spectral purity \cite{laatikainen2023reduction}. In the former case Mandel's condition is fulfilled at some special points, whereas in the latter it is fulfilled across the whole wavefront.

\begin{figure}
    \centering
    \includegraphics[width=\columnwidth]{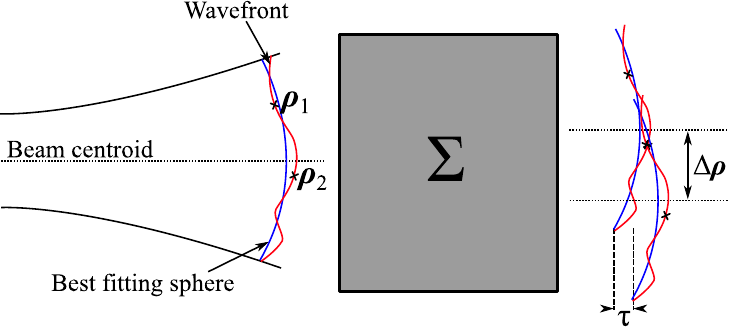}
    \caption{Beam of light is incident from the left onto a measurement device $\Sigma$. The effect of the device is to split the field into two replicas, with a controllable amount of lateral and longitudinal shift ($\Delta\boldsymbol{\rho}$ and $\tau$, respectively). This allows one to measure the spectrum of any desired superposition at the output of the device.}
    \label{fig:a}
\end{figure}

In the following, we shall consider ideal field superpositions, which may be generated with the use of a wavefront-folding, or -shearing interferometer \cite{turunen2022measurement}, as depicted schematically in Fig.~\ref{fig:a}. The superposition at position $\boldsymbol{R}$ of two fields emanating from $\boldsymbol{\rho}_1$ and $\boldsymbol{\rho}_2$ can be written in the general form
\begin{align}
\label{eq38}
    E(\boldsymbol{R};\omega) = E(\boldsymbol{\rho}_1;\omega) + E(\boldsymbol{\rho}_2; \omega)\exp(i\omega\tau), 
\end{align}
where $\tau$ is a variable time difference. Substituting the field from Eq. (\ref{eq38}) into Eq. (\ref{eq39}), we obtain the CSD at the plane of interference as 
\begin{align}
\label{eq44}
    W(\boldsymbol{R},\boldsymbol{R};\omega_1,\omega_2) & = W(\boldsymbol{\rho}_1,\boldsymbol{\rho}_1;\omega_1,\omega_2) \nonumber \\
    & + W(\boldsymbol{\rho}_1,\boldsymbol{\rho}_2;\omega_1,\omega_2)\exp(i\omega_2\tau) \nonumber \\
    & + W(\boldsymbol{\rho}_2,\boldsymbol{\rho}_1;\omega_1,\omega_2) \exp(-i\omega_1\tau) \nonumber \\
    & + W(\boldsymbol{\rho}_2,\boldsymbol{\rho}_2; \omega_1,\omega_2)\exp[{-i(\omega_1-\omega_2)\tau}].   
\end{align}
By setting $\omega_1 = \omega_2 = \omega$, the spectral density at point $\boldsymbol{R}$ is
\begin{align}
\label{eq45}
    S(\boldsymbol{R} ;\omega) & = S(\boldsymbol{\rho}_1;\omega) + S(\boldsymbol{\rho}_2; \omega) + 2\sqrt{S(\boldsymbol{\rho}_1;\omega)S(\boldsymbol{\rho}_2; \omega)} \nonumber\\
    & \quad \times \left|\mu(\boldsymbol{\rho}_1,\boldsymbol{\rho}_2; \omega,\omega)\right| \cos\left[\phi(\boldsymbol{\rho}_1,\boldsymbol{\rho}_2; \omega,\omega) + \omega\tau \right],
\end{align}
where $\phi(\boldsymbol{\rho}_1,\boldsymbol{\rho}_2; \omega,\omega)$ is the phase of the complex degree of coherence. Let us now consider the special case of complete spatial coherence, for which $\left|\mu(\boldsymbol{\rho}_1,\boldsymbol{\rho}_2; \omega,\omega)\right| = 1$ for all spatial coordinates. We can rearrange Eq.~(\ref{eq45}) to obtain
\begin{align}
    \frac{S(\boldsymbol{R} ;\omega) - S(\boldsymbol{\rho}_1;\omega) - S(\boldsymbol{\rho}_2; \omega)}{2\sqrt{S(\boldsymbol{\rho}_1;\omega)S(\boldsymbol{\rho}_2; \omega)}} =
    \cos\left[\phi(\boldsymbol{\rho}_1,\boldsymbol{\rho}_2; \omega,\omega) + \omega\tau \right],
\end{align}
which is a convenient form to discuss the effects of cross-spectral purity.

Let us next assume that the condition of Eq.~(\ref{condition}) holds, so that we may write
\begin{align}
\label{phase}
    \frac{D(\boldsymbol{R},\boldsymbol{\rho}_1) - 1 - C(\boldsymbol{\rho}_1,\boldsymbol{\rho}_2)}{2\sqrt{C(\boldsymbol{\rho}_1,\boldsymbol{\rho}_2)}} = \cos\left[\phi(\boldsymbol{\rho}_1,\boldsymbol{\rho}_2; \omega,\omega) + \omega\tau \right].
\end{align}
The first equality in Eq.~(\ref{condition}) states that the normalized spectra at the two points $\boldsymbol{\rho}_1$ and $\boldsymbol{\rho}_2$ are equal. If this condition holds across the whole wavefront, then the average energy spectrum (i.e. average spectral amplitude) does not depend on spatial position. The second equality states that the normalized spectrum at the superposition of the fields belonging to $\boldsymbol{\rho}_1$ and $\boldsymbol{\rho}_2$ are equal. By inspecting Eq.~(\ref{phase}) we note that for the second equality in Eq.~(\ref{condition}) to hold, the right hand side must not depend on frequency. {This was solved in Ref.~\cite{koivurova2019cross} by assuming a spatiotemporally separable correlation function, establishing a relation from spatiotemporal separability to cross-spectral purity. Here, we are concerned with the reverse, and to show that it is in fact a two-way relation under certain circumstances.} Let us next consider the structure of the phase $\phi(\boldsymbol{\rho}_1,\boldsymbol{\rho}_2; \omega,\omega)$ in detail. 

The phase of the complex degree of coherence is directly related to the phase of the electric field, and in the case of complete coherence it can be written as $\phi(\boldsymbol{\rho}_1,\boldsymbol{\rho}_2; \omega,\omega) = \varphi(\boldsymbol{\rho}_2;\omega) - \varphi(\boldsymbol{\rho}_1;\omega)$. The phase of the electric field can contain three different types of contributions: i) an absolute phase term $\varphi_a$ which does not depend on spatial position or frequency, ii) a phase component which depends on either spatial position $\varphi_s(\boldsymbol{\rho})$, or frequency $\varphi_f(\omega)$ such that both can be present at the same time, and iii) a term which depends on both space and frequency $\varphi_{sf}(\boldsymbol{\rho};\omega)$. 

The absolute phase term has no detectable consequences in an interferometric experiment, and therefore can be neglected. Similarly, phase terms which depend only on frequency are not detectable, since they appear in the form $\Delta\varphi_f(\omega_1,\omega_2) = \varphi_f(\omega_1) - \varphi_f(\omega_2)$ and we have chosen $\omega_1 = \omega_2 = \omega$, in which case $\Delta\varphi_f(\omega,\omega) = 0$. 

On the other hand, the space dependent phase is detectable, since it has the form $\Delta\varphi_s(\boldsymbol{\rho}_1,\boldsymbol{\rho}_2) = \varphi_s(\boldsymbol{\rho}_1) - \varphi_s(\boldsymbol{\rho}_2)$ and now $\boldsymbol{\rho}_1$ is not necessarily equal to $\boldsymbol{\rho}_2$. The spatial phase term imposes a uniform and constant phase shift onto all frequency components, depending on the spatial position. As such, it is similar to an absolute phase component, and it will only shift the fringes. Therefore, while it can be detected (at least in principle) its effect is minor.

The last phase component has the form $\Delta\varphi_{sf}(\boldsymbol{\rho}_1,\boldsymbol{\rho}_2;\omega) = \varphi_{sf}(\boldsymbol{\rho}_1;\omega) - \varphi_{sf}(\boldsymbol{\rho}_2;\omega)$, which imposes a frequency dependent phase onto the field as a function of position. If we expand the spatiospectral phase term into a series with respect to frequency, we see that it can be written in the form
\begin{align}\label{expansion}
    \varphi_{sf}(\boldsymbol{\rho};\omega) = \varphi_0(\boldsymbol{\rho};\omega_0) + \varphi_1(\boldsymbol{\rho};\omega_0)\omega + \varphi_2(\boldsymbol{\rho};\omega_0)\omega^2 + ...
\end{align}
where $\omega_0$ is some reference frequency, such as the center frequency of the spectrum. We can further write
\begin{align}
    \Delta\varphi_{sf}(\boldsymbol{\rho}_1,\boldsymbol{\rho}_2;\omega) & = \varphi_{sf}(\boldsymbol{\rho}_1;\omega) - \varphi_{sf}(\boldsymbol{\rho}_2;\omega) \\ \nonumber
    & = \Delta\varphi_0(\boldsymbol{\rho}_1,\boldsymbol{\rho}_2;\omega_0) + \Delta\varphi_1(\boldsymbol{\rho}_1,\boldsymbol{\rho}_2;\omega_0)\omega + ...
\end{align}
and each term in the expansion has a definite role. The zeroth order term again only causes a shift in the interference fringes. In fact, we can identify $\Delta\varphi_{sf}(\boldsymbol{\rho}_1,\boldsymbol{\rho}_2;\omega_0) \equiv \Delta\varphi_s(\boldsymbol{\rho}_1,\boldsymbol{\rho}_2)$, since the zeroth order term does not explicitly depend on frequency (the implicit dependence on $\omega_0$ is due to the series expansion). 

The linear first order term corresponds to a shift in the temporal domain, while leaving the pulse shape intact. In other words, if the wavefront is shaped such that the electric field around $\boldsymbol{\rho}_1$ precedes (or lacks behind) the electric field around $\boldsymbol{\rho}_2$, then we see sinusoidal interference fringes depending on the magnitude of the delay. Sources of second order spatiospectral phase include reflection from (or transmission through) a curved or rough surface, as well as free space propagation which induces a spherical phase. From here we see that the role of $\tau$ in a measurement of cross-spectral purity is to compensate for the possible timing mismatch between points $\boldsymbol{\rho}_1$ and $\boldsymbol{\rho}_2$ (as depicted in Fig.~\ref{fig:a}). This can be achieved by choosing
\begin{align}
    \tau = -\Delta\varphi_1(\boldsymbol{\rho}_1,\boldsymbol{\rho}_2;\omega_0).
\end{align}
We note that the measurement setup induced delay $\tau$ is variable, so that if one laterally scans across the whole wavefront, then it is always possible to find a delay that causes the linear fringes to vanish at a given point. On the other hand, the variable delay can be used for wavefront sensing. Additionally, one can use $\tau$ to induce interference fringes and measure the spatial phase term $\Delta\varphi_s(\boldsymbol{\rho}_1,\boldsymbol{\rho}_2)$ if necessary. We note that in general, the wavefront may have a strongly varying shape. Nonetheless, it is possible to correct the shape of the wavefront with e.g. the use of a deformable mirror, and then perform cross-spectral purity measurements at the focal plane of a lens to ensure that the wavefront is planar. This makes the measurements vastly simpler, since one would not need to vary the delay $\tau$ for all different spatial points.

All the higher orders in Eq.~(\ref{expansion}) contribute to the shape of the pulse, such that moving along the wavefront one may see many different pulse shapes. A realistic example of fields with spatially varying pulse shapes are isodiffracting pulsed beams, which are produced in spherical mirror cavities \cite{koivurova2018partially}. Incidentally, these higher orders will cause nonlinear spectral interference fringes at the plane of superposition, thus clashing with the condition for cross-spectral purity. On the other hand, it is entirely possible that there are again some special points where the higher order contributions exactly cancel, but this does not indicate that the spectral phase does not depend on position. Only in the case of global purity (i.e. no observed fringes for all possible combinations of $\boldsymbol{\rho_1}$ and $\boldsymbol{\rho_2}$) that is the case.

Therefore, we have shown that in the case of a completely coherent field that is globally cross-spectrally pure, the second equality in Eq.~(\ref{condition}) ensures that the spectral phase does not depend on spatial position. By linearity of Fourier transform, if a completely coherent field is cross-spectrally pure at the plane of measurement, then it is also spatiotemporally separable. This means that there is a two-way relation between global cross-spectral purity and spatiotemporal separability.

Conversely, in the case of complete spatial incoherence -- $\left|\mu(\boldsymbol{\rho}_1,\boldsymbol{\rho}_2; \omega,\omega)\right| = 0$ for all $\boldsymbol{\rho}_2 \neq \boldsymbol{\rho}_1$ -- it is trivial to show that global cross-spectral purity does not indicate separability of the field. Note that we do not restrict spectral correlations. In this case, the spectral interference law of Eq.~(\ref{eq45}) becomes
\begin{align}
\label{part1}
    S(\boldsymbol{R} ;\omega) = S(\boldsymbol{\rho}_1;\omega) + S(\boldsymbol{\rho}_2; \omega),
\end{align}
for all $\boldsymbol{\rho}_2 \neq \boldsymbol{\rho}_1$, and
\begin{align}
\label{part2}
    S(\boldsymbol{R} ;\omega)  = 2S(\boldsymbol{\rho}_1;\omega) \left[1 + \cos\left(\omega\tau \right) \right],
\end{align}
when $\boldsymbol{\rho}_2 = \boldsymbol{\rho}_1$. Assuming that Eq.~(\ref{condition}) holds globally and we choose $\tau=0$, we can write Eqs.~(\ref{part1}) and (\ref{part2}) together as
\begin{align}
    D(\boldsymbol{R},\boldsymbol{\rho}_1) = \left[ 1 + C(\boldsymbol{\rho}_1,\boldsymbol{\rho}_2) \right].
\end{align}
This expression holds for all combinations of $\boldsymbol{\rho}_1$ and $\boldsymbol{\rho}_2$, and it does not contain any frequency dependent contributions, since the lack of spatial correlations hides them. Therefore, in the case of a completely spatially incoherent field, we have a one way relation between global cross-spectral purity and spatiotemporal separability of the field. {However, the spatiotemporal \textit{correlation function} of a cross-spectrally pure field always obeys the reduction formula \cite{laatikainen2023reduction}, 
\begin{align}
    \bar{\gamma}(\boldsymbol{\rho}_1,\boldsymbol{\rho}_2;\Delta t - \tau) = \bar{\gamma}(\boldsymbol{\rho}_1,\boldsymbol{\rho}_2; -\tau)\bar{\gamma}(\boldsymbol{\rho}_1,\boldsymbol{\rho}_1;\Delta t) 
\end{align}
which is a manifestation of spatiotemporal separability of partially coherent fields. In essence, the reduction formula states that the full correlation function can be expressed as a product of a spatial and a temporal term. This is fully compatible with the transformations of correlation functions of iso-diffracting fields to spatiotemporally separable ones discussed in Ref.~\cite{laatikainen2022spectral}, which are applicable even in the case of complete spatial incoherence. Thus, it is possible to have a completely spatially incoherent field that has a correlation function that separates into spatial and temporal contributions.}

The case of partial coherence is more involved, since now the absolute value of the complex degree of spectral coherence has a value between 0 and 1, depending on the three coordinates $\boldsymbol{\rho}_1,\boldsymbol{\rho}_2,\omega$. Moreover, the phase $\phi(\boldsymbol{\rho}_1,\boldsymbol{\rho}_2;\omega,\omega)$ no longer directly corresponds to the phase of the field, but rather, an average over many realizations. Nonetheless, it is conceivable that similar reasoning as in the completely coherent case could be used to establish whether a cross-spectrally pure partially coherent field is separable or not.

However, the analysis of completely incoherent fields gives us an important clue: if there is a combination of coordinates for which $\mu(\boldsymbol{\rho}_1,\boldsymbol{\rho}_2;\omega,\omega) = 0$, then one may have problems establishing whether the field is separable or not. If the correlation function has an extend area of incoherence, then it becomes impossible to establish whether a field is spatiotemporally separable or not, even if it is cross-spectrally pure. Therefore, extended areas of incoherence in the correlation function inhibit our ability to establish spatiotemporal separability, and the conclusions one may draw from measurements of cross-spectral purity depend strongly on the functional form of the coherence function.

{Hence, global cross-spectral purity does not automatically ensure spatiotemporal separability of the \textit{field} in the case of partial coherence. We note again that the reduction formula applies \cite{laatikainen2023reduction}, and it is possible to transform partially coherent iso-diffracting fields into spatiotemporally separable ones that fulfill Mandel's condition globally \cite{laatikainen2022spectral}. In other words, there are partially coherent fields which have a spatiotemporally separable correlation function, and thus are also globally cross-spectrally pure. Therefore, we have shown that the concept of cross-spectral purity generalizes spatiotemporal separability to fields of any state of coherence, through their correlation functions. This is the main result of the present work.}

As a last note, we consider how to measure cross-spectral purity. Let us assume that the wavefront is planar (i.e. corrected wavefront and a measurement at the focal plane of a lens), so that we do not need to vary $\tau$. To measure cross-spectral purity, one needs to first measure the position dependent spectrum with e.g. an imaging spectrometer, and then measure the position dependent spectrum of a superposition of the form of Eq.~(\ref{eq38}). Such superpositions can be easily performed with the use of a wavefront-folding, or wavefront-shearing interferometer \cite{turunen2022measurement}. The beams coming from both arms of the interferometer have to be set parallel to each other, in order to avoid spatial fringes. At the output of the interferometer an imaging spectrometer measures the position dependent spectrum. The two beams must be scanned over each other, to access all possible combinations of $\boldsymbol{\rho}_1$ and $\boldsymbol{\rho}_2$. Such experiments are conceptually simple to perform, but will produce a large amount of data (a three dimensional data cube for each beam overlap position). {It needs to be noted that the measurements discussed here yield information on relative spatiospectral couplings, and to obtain absolute couplings one needs to supplement the data with a pulse measurement. This is similar to methods such as TERMITES or INSIGHT \cite{Miranda:14,Pariente2016,Borot:18}} 

In conclusion, we have shown that global cross-spectral purity indicates spatiotemporal separability of the field when the field is completely spatially coherent. That is, in the case of complete spatial coherence, the first equality of Eq.~(\ref{condition}) ensures that the spectral amplitude does not depend on spatial position, whereas the second equality ensures that the spectral phase does not depend on spatial position. {The case of partial spatial coherence is strongly affected by the functional form of the correlations, while in the case of complete spatial incoherence one cannot establish separability of the field from measurements of cross-spectral purity. However, due to the reduction formula, cross-spectral purity generalizes the concept of spatiotemporal separability of fields to spatiotemporal separability of correlation functions. This is of fundamental importance in both theory and experiments. Since the notion of cross-spectral purity has recently been defined for nonstationary vector fields \cite{joshi2021cross,laatikainen2023cross,joshi2024cross}, the present findings can be straightforwardly extended to take into account the vectorial nature of light as well.}

\footnotesize
\section*{Funding}
Research Council of Finland (346518)

\section*{Disclosures}
 The authors declare no conflicts of interest.
 
\section*{Data Availability}
Data regarding this work is currently not in the public domain but may be obtained from the authors on reasonable request.

\clearpage

\bibliography{references}
\bibstyle{plain}
\end{document}